\documentclass[mathleft
]{an}
\usepackage{graphicx}
\usepackage{times}
\usepackage{color}

\begin{document}

\Pagespan{1}{}
\Yearpublication{2010}%
\Yearsubmission{}%
\Month{}%
\Volume{}%
\Issue{}%

\title{Size matters}

\author{J.~C.\ del Toro Iniesta\inst{1}\fnmsep\thanks{\email{jti@iaa.es}\newline}
\and  D. Orozco Su\'arez\inst{2,1}
}
\titlerunning{Size matters}
\authorrunning{Del Toro Iniesta \& Orozco Su\'arez}
\institute{Instituto de Astrof\'{\i}sica de Andaluc\'{\i}a (CSIC),
Apdo.\ de Correos 3004, 18080 Granada, Spain
\and 
  National Astronomical Observatory of Japan,
2-21-1 Osawa, Mitaka, Tokyo 181-8588, Japan}

\received{}
\accepted{}
\publonline{later}


\abstract{The new generation of ground-based, large-aperture solar telescopes promises to significantly increase our capabilities to understand the many basic phenomena taking place in the Sun at all atmospheric layers and how they relate to each other. A (non-exaustive) summary of the main scientific arguments we have to pursue these impressive technological goals is presented. We illustrate how imaging, polarimetry, and spectroscopy can benefit from the new telescopes and how several wavelength bands should be observed to study the atmospheric coupling from the upper convection zone all the way to the corona. The particular science case of sunspot penumbrae is barely discussed as a specific example.}

\maketitle

\section{Introduction}

Understanding the plasma physical processes occurring at the Sun at all scales has always progressed in parallel to the development and construction of new solar telescopes, post-focus instrumentation, and data analysis techniques. In a large measure because of the new facilities, our knowledge has increased significantly in recent years, but still much effort is needed to address critical questions. The plasma and magnetic field interactions that take place in the solar atmosphere involve characteristic spatial and temporal scales that are too small to be fully resolved with current solar facilities, and leave tracks in the spectrum of polarized light that can hardly be detected with current instrumentation. These problems can be alleviated by increasing our photon flux budget capabilities. Therefore, we need larger solar telescopes. Indeed, a number of them are now being designed and constructed. The two biggest projected telescopes are the EST (European Solar Telescope) and the ATST (Advanced Technology Solar Telescope), each having a photon collecting area larger than the sum of all other currently available and planned areas. Both, thus, constitute a solid promise of a significant qualitative advancement for the solar physics community. In some provocative way, then, we can say as the title of this paper reads that ``size matters''.

Summarizing the main reasons that justify the strong effort of building 4-meter class solar telescopes or, in other words, prospecting for the likely scientific advances to be reached with them is a difficult endeavor. As a matter of fact, a great deal of information on that topic can already be found in the two excellent science requirement documents of the mentioned projects and on, e.g., Keil et al. (2001, 2003, 2004, 2009), Keller et al. (2002), Rimmele et al. (2003, 2005), or Collados (2008). An exhaustive search cannot be expected in this paper. Rather, we try to highlight some special topics and to give some specific examples that are very important according to our personal point of view.

\begin{figure*}[!t]
\centering
\resizebox{\hsize}{!}{\includegraphics{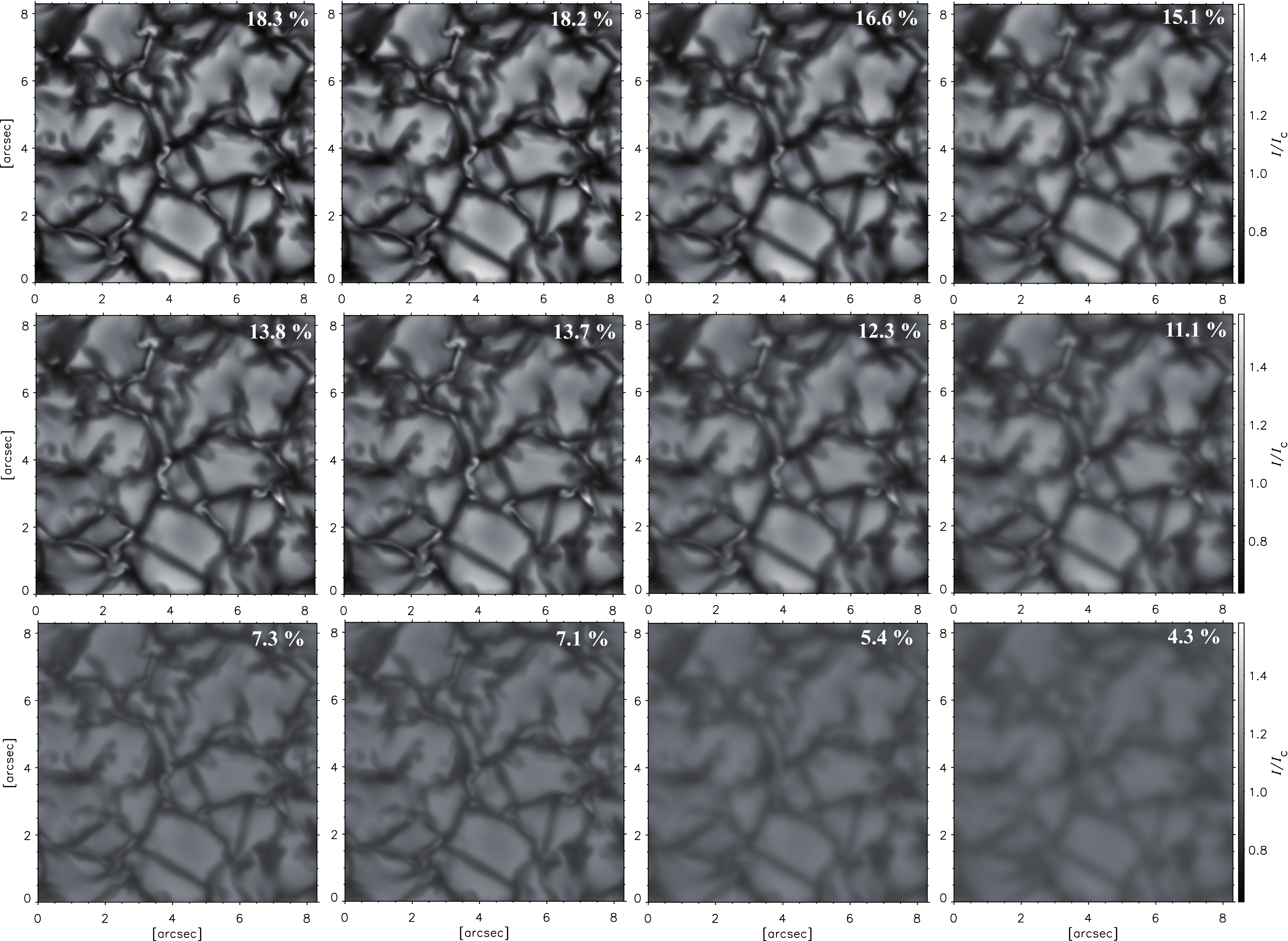}}
\caption{Simulated continuum intensity maps corresponding to telescopes with apertures of 4 (first two columns), 1.5, and 1 m, representative (from left toward right) of the ATST, EST (4 m with a central obscuration), Gregor, and Sunrise and SST, respectively, all considered without atmospheric effects. The three rows correspond to three different wavelengths, namely, 525, 630, and 1560~nm from top to bottom. Text inserts show the corresponding rms contrasts. The simulation snapshot has an average vertical field of 10~G. Only diffraction effects are considered.} 
\label{fig:continuos} 
\end{figure*} 

\section{Main drivers for large-aperture telescopes}
\label{sec:labels}

\subsection{Spatial resolution} 
\label{sec:resolution}

Solar features and phenomena can be observed at a vast variety of length scales, the smallest in the photosphere being probably of the order of kilometers or less (e.g., de Wijn et al.\ 2009). Our current facilities are still far from resolving these tiny details, most of which are presumably close to the mean free path of photons (a few tens of a km). The first, straightforward approach implies a significant increase in telescope apertures. The direct influence of the size is clearly understood after a glance to Fig.\ \ref{fig:continuos}, where synthetic images are shown that have been obtained through radiative transfer calculations with the SIR code (Ruiz Cobo \& Del Toro Iniesta, 1992) on MHD numerical simulations generated with the MURAM code (V\"ogler et al.\ 2005) for a solar zone with an average vertical magnetic field of 10~G. Just the diffraction effects are taken into account for producing the images. The aperture diameters range from 4 m (ATST and EST), to 1.5 m (Gregor), and 1 m (Sunrise and SST). At given wavelengths differences in contrast are significant. For example, values go from 18.3 \% with 4-m telescopes to 15.1 \% with 1-m ones at 525 nm. The slight differences between the two 4-m telescopes are due to the EST central obscuration. Note that no atmospheric seeing is included in the synthetic images. Hence, these images represent theoretical limits for the corresponding instrumentation that, everybody knows, are hardly reachable on ground. The most likely option to approximate these limits is adaptive optics (AO; e.g., Berkefeld et al. 2002, Scharmer et al. 2003, Rimmele et al. 2005, and references therein) or post-facto restoring techniques like MOMFBD (multi-object, multi-frame, blind deconvolution; van Noort et al. 2005) or phase diversity (Paxman et al. 1996), that can even be useful for space- or balloon-borne observations when corrections for residual jittering or other motions are necessary. Such an approximation has only been possible for individual images or for not-long-enough data series. MHD simulations have shown, however, that many magnetic processes take place at such small scales. But even if the ideal resolution is not exactly reached, we need to continue exploring whether or not smaller and smaller scales seem to exist. Such analyses will, in turn, feed back the MHD modeling. Another clear feature in Fig.\ \ref{fig:continuos} is the importance of the observing wavelength. As soon as we go to the infrared (IR), contrast deteriorates dramatically. We can be interested in the increased diagnostic potential of the IR wavelengths. Thus, if we aim at spatial resolutions in these wavelengths similar to those currently reached in the visible with telescopes smaller than 1 m, we necessarily have to pursue the use of large-aperture telescopes.

\begin{figure}[!t]
\centering 
\resizebox{\hsize}{!}{\includegraphics{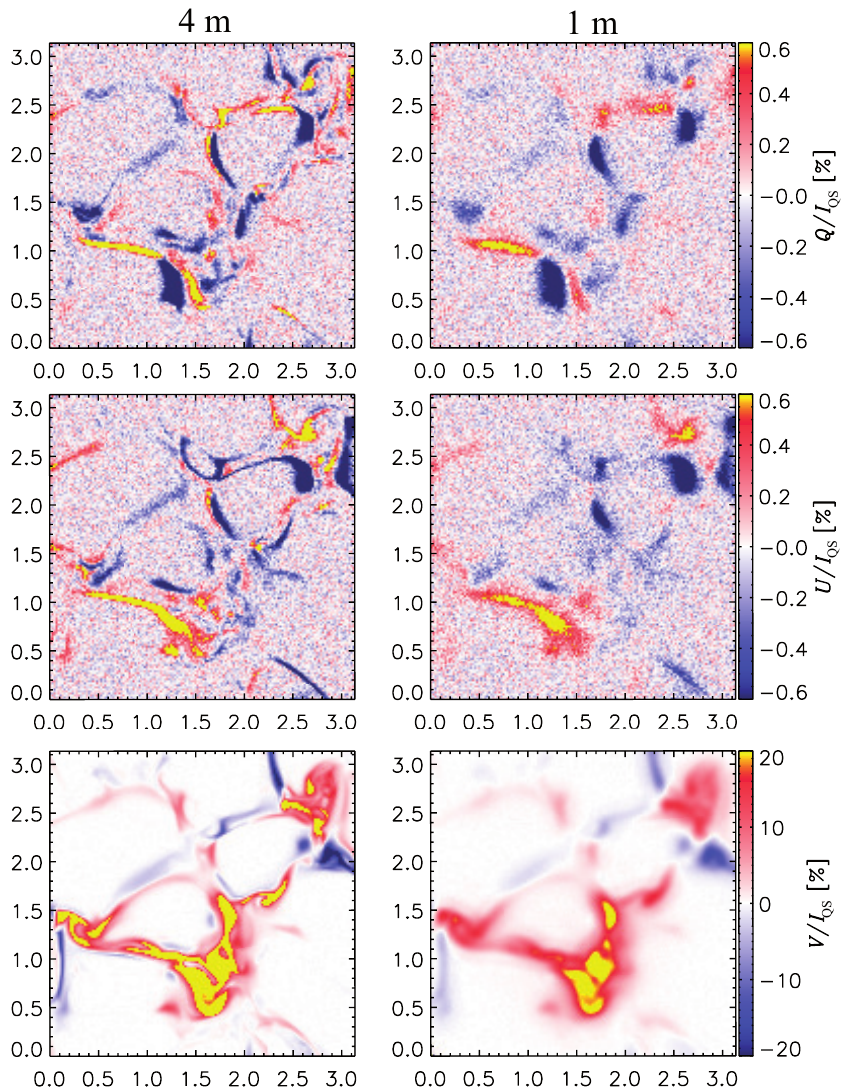}}
\caption{Monochromatic Stokes $Q$ (top), $U$ (middle), and $V$ (bottom) images at +7.7 pm of the central wavelength of the Fe~{\small I} line at 525.02~nm. Left and right columns show the images through a diffraction-limited telescope of 4 and 1 m, respectively. We have added noise at the level of $10^{-3}\, I_\mathrm{QS}$, where $I_\mathrm{QS}$ is the average continuum of the quiet Sun. Grid units are in arcsec.}  
\label{fig:mapaspol} 
\end{figure}

Spatial resolution is also paramount for spectropolarim-etry in order to fully characterize the many small-scale magnetic fields that are known to populate the solar atmosphere. The larger the telescope aperture, the less distorted the polarization maps. Figure \ref{fig:mapaspol} shows an example of Stokes $Q$, $U$, and $V$ polarization maps as seen  by a 4-m and a 1-m telescopes (left and right columns, respectively). Signals at a fixed wavelength (+ 7.7 pm far from line center) of the Fe~{\sc i} line at 525.02~nm are considered. Right panels show less polarization signals and magnetic structuring than the left panels. Besides, tiny details scape from detection in the 1-m maps. In view of this, should we still speak of filling fractions when trying to model the (currently) unresolved magnetic structures when observed with larger telescopes? Recent advances in spatial resolution with accurate instrumentation like the solar optical telescope (Tsuneta et al. 2008) aboard {\em Hinode} (Kosugi et al. 2007) are bringing about a new paradigm of the quiet Sun: it has been unveiled to be covered by tiny, mostly horizontal, magnetic structures (Lites 2007, 2008). From 0 (we had not detected them properly yet) we now estimate a filling factor about $0.2 - 0.45$ (Orozco Su\'a\-rez et al. 2007) of the resolution element. Is this fraction still growing (and decreasing in other places) with the increasing size of telescopes? We urgently need an answer to this question: should the answer be negative at a given resolution we would conclude on having a homogeneous distribution of magnetic features; a positive answer would entail the conclusion of a heterogeneous distribution. Elucidating between these two cases has important consequences about the nature (homogeneously turbulent according to some authors) of the internetwork magnetic fields and about the dynamo action behind them.

\begin{figure}[!t]
\centering 
\resizebox{\hsize}{!}{\includegraphics{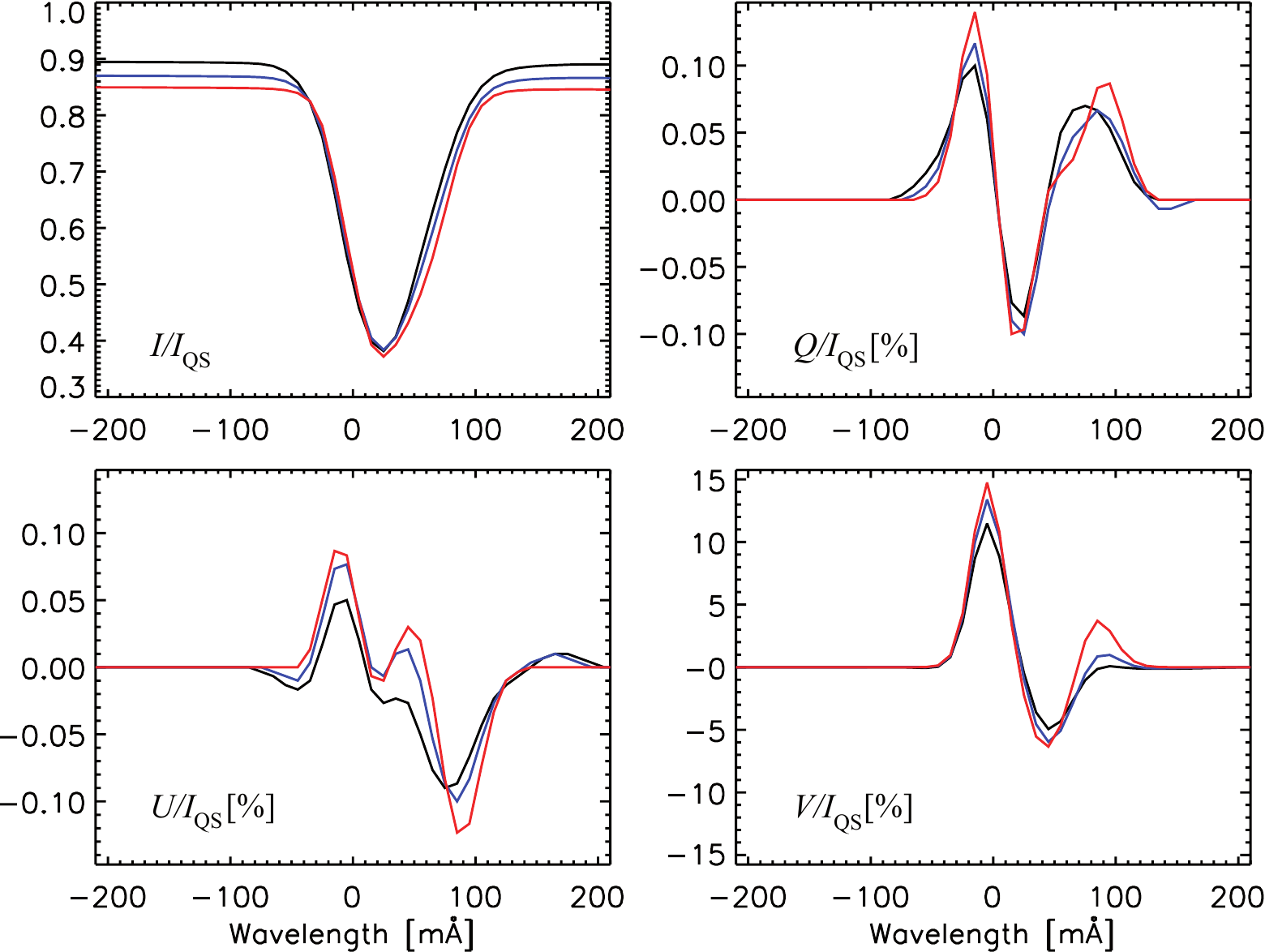}} 
\caption{Synthetic Stokes profiles of the Fe~{\small I} line at 525.02~nm as seen by a 4-m (red), a 1.5-m (blue), and 1-m telescope (black).} 
\label{fig:spectrum} 
\end{figure}

Spatial resolution is not only important for imaging and polarimetry. Spectroscopy is also enriched by larger apertures. Although very simple and easy to understand, little attention has received so far this effect in the literature (Oroz\-co Su\'arez et al. 2010). The smaller the resolution element, the larger the details in the spectrum. Figure\ \ref{fig:spectrum} shows the effect of telescope diffraction on the Stokes profiles of the Fe~{\sc i} line at 525.02 nm as emerging from one point of the MHD simulations. Red lines represent the original resolution of the numerical simulations (almost equal to that of 4-m, diffraction-limited telescope), in blue are the profiles as seen by a 1.5-m telescope, and the observed spectrum by a 1-m telescope is seen in black.

\subsection{Photon budget} 
\label{sec:photon}

We measure nothing but light and, therefore, our measurements rely upon photometric accuracy, $\delta I / I$. This quantity represents the (relative) smallest detectable signal and is inversely proportional to the signal-to-noise ratio: $S/N = 1/(\delta I / I)$. If $p$ represents the degree of polarization (no matter linear, circular, or total), a simple algebra leads us to $\delta p /p \, \mbox{\raisebox{-.6ex}{$\stackrel{ <}{\sim}$}} \, \sqrt{1 + 1/p^{2}}/(S/N)$.\footnote{The asymptotic behavior of $\delta p/p$ for $p=0$ simply reflects the fact that polarimetric accuracy worsens for very small degrees of polarization.} This means that if we want to reliably measure the faint polarization signals that are detected of the order of $10^{-3} - 10^{-4}\, I_{\rm QS}$ with low spatial and temporal resolution (e.g., L\'opez Ariste \& Casini 2003, Stenflo 2006, Trujillo Bueno 2009), then we need $S/N$ values well above $10^{3} - 10^{4}$. Enhancing the $S/N$ at a given bandwidth is only possible by increasing the telescope aperture or the exposure time. Since we cannot afford the latter option because the extremely dynamic nature of most of the phenomena under analysis, it seems that we should pursue the construction of larger telescopes. This will especially be true in the chromosphere where the rapid evolution of features escapes detection with long exposure times even in total intensity in the continuum. 

The solar surface shows very dynamic phenomena. Evershed (often supersonic) flows and moving magnetic features in sunspots, convective collapse of flux tubes, superso\-nic horizontal flows in granules, moving spicules and other features in the chromosphere, high-frequency prominence oscillations, and magnetic reconnection are some examples. Such processes occur at short time scales that need be resolved for a meaningful analysis. Even more, as soon as we increase the resolution, we expect to detect new small-scale magnetic structures whose timescales would demand even faster-cadence observations. Therefore, a better photon budget will improve the temporal resolution and allow better coverage for evolutionary studies. 

\subsection{Wavelength coverage}

One of the most challenging goals for the near future is to understand and quantify the magnetic coupling of the whole atmosphere from the upper convection zone through the corona. We need to study how energy is transported and dissipated and what is the role of magnetic fields in this process. Multi-layer probing can only be possible by observing several wavelength regions at the same time with the same telescope and the most efficient way to do that is to direct different bands to different instruments. The rapid decrease of the Planck function towards the ultraviolet and the infrared demands large apertures in order to reach the best observing conditions in these two regions of the spectrum. They are crucial for chromospheric and coronal studies. Nevertheless, less ambitious objectives can also benefit from a broad wavelength coverage as already demonstrated by, for example, the simultaneous visible and IR observations by Cabrera Solana et al. (2006) that evidenced how moving magnetic features have their origin in the Evershed effect, or by Mart\'{\i}nez Gonz\'alez et al. (2008) that found hG magnetic fields in the internetwork. The increased information of the two wavelength bands, improves the reliability of their conclusions.

\subsection{Coronography}

Measuring coronal magnetic fields is a goal (and a challenge) in itself. The physical phenomena taking place in the chromosphere and corona are better observed through polarization by means of the Zeeman effect in emission lines or of the Hanle measurements in scattered radiation. The low degrees of polarization on very low brightness structures makes current discoveries painful. Examples are, e.g., the detection of an extended near-Sun He~{\sc i} cloud by Kuhn et al.\ (2007), the measurement of a 4 G magnetic flux density 100 arcsec above an active region by (Lin et al.\ 2004), or the detection of Alfv\'en waves in intensity, line-of-sight velocity and linear polarization images by Tomczyk (2007). Large-aperture telescopes are needed in order not to be limited by photon noise. The ATST specific design for coronography is expected to further help. 

\section{A science case: the sunspot penumbra}
\label{sec:penumbra}

\begin{figure}[!t]
\centering 
\resizebox{\hsize}{!}{\includegraphics{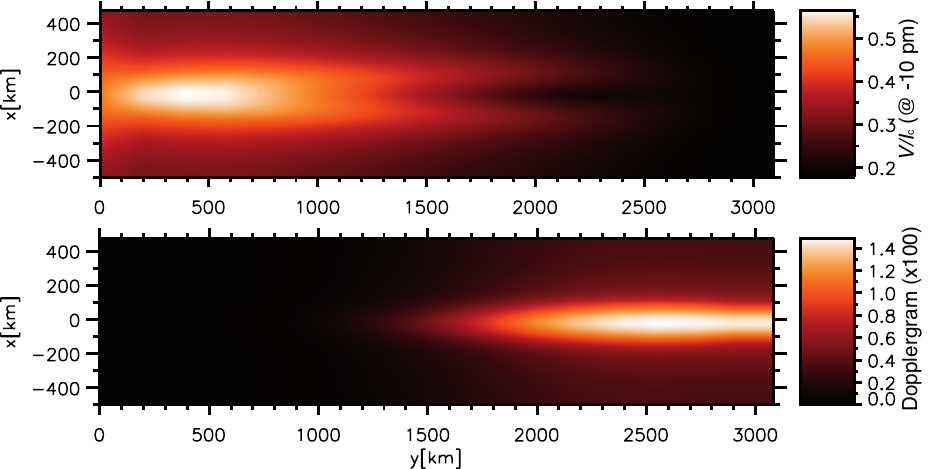}} 
\caption{Magnetogram and Dopplergram of the dark-cored, bright penumbral filament model by Ruiz Cobo \& Bellot Rubio (2008). Both diagrams are dimensionless. The first is equal to the Stokes $V/I_{\rm c}$ signal at $-10$ pm from the line center; the second is the normalized difference of Stokes $I$ at $\mp 15$ pm. (Receding velocities are positive.)}
\label{fig:filamento} 
\end{figure}

We do not want to end this contribution without commenting on a specific scientific case that is very important to our personal interests and that will clearly benefit from the advancement in instrumentation. Sunspot penumbrae are am\-ong the oldest observable solar features that still puzzle us. Two are in our opinion the most important discoveries of the last fifteen years related to the sunspot penumbra, namely, the discovery of an Evershed downward mass flux at its outer periphery by Westendorp Plaza et al. (1997) and the discovery of dark cores along bright penumbral filaments by Scharmer et al. (2002). These and other observational facts that have been gathered by the whole community need to be explained by any model trying to give account of the nature of the penumbra. A list of the most important ones include: 1) the penumbra is bright; 2) the Evershed flow takes place preferentially in the dark cores (Bellot Rubio et al. 2005); 3) it returns to the surface at the middle penumbra and beyond (e.g. Ichimoto et al. 2007); 4) it is often supersonic (e.g., Wiehr 1995, Del Toro Iniesta et al. 2001, Bellot Rubio et al. 2004); 5) it is magnetized (e.g. S\'anchez Almeida \& Lites 1992, Mart\'inez Pillet 2000, Westendorp Plaza et al. 2001a,b); 6) it is associated to the weakest and more inclined magnetic fields (see Bellot Rubio 2009 and Tritschler 2009 for reviews); and 7) it continues beyond the outer penumbral border, often as moving magnetic features (Sainz Dalda \& Mar\-t\'{\i}nez Pillet 2005, Cabrera Solana et al. 2006, Ravindra 2006, Ku\-bo et al. 2007). Two are as well the competing theoretical models for explaining the nature of the penumbra: the uncombed model by Solanki and Montavon (1993) that, after the theoretical calculations of Ruiz Cobo \& Bellot Rubio (2008; see a magnetogram and a Dopplergram resulting from these calculations in Fig.\ \ref{fig:filamento}) of a hot plasma flowing through the dark cores of penumbral filaments explains most of the above observational facts; and the gappy penumbra by Spruit \& Scharmer (2006) and Scharmer \& Spruit (2006) where field-free gaps protrude the magnetic penumbra carrying energy from below and, thus, heating the pen\-umbra through overturning convection.  This mechanism of overturning convection is far from being firmly established by observations since no velocities are detected along the borders of penumbral filaments. Spatial resolutions much better than 0\farcs 1 are certainly needed to settle the debate.

\section{Concluding remarks}

After having carried out our personal review, we must agree with the title of the paper: size of telescopes certainly matters. But the observing wavelength and coverage matter too, and the seeing conditions for the observation, and the quality of the AO system, and, although not very much discussed in here, the performance of instruments and of the analysis techniques is very important to rely upon the results. Moreover, our prejudices based on the current paradigm are sometimes determinant for reaching one conclusion or another. In summary, the ideal situation would be such that a large-aperture telescope feeds several highly efficient, state-of-the-art instruments working at several wavelength bands, at an extremely good site where seeing conditions are nevertheless improved with a very powerful AO system. The data are then analyzed with sophisticated inversion analysis techniques that may take several different scenarios (set of hypotheses) into account in order to discriminate among prejudices of the different researchers. This can only be accomplished after lots of professional imagination and expertise are put to the service of the community. Therefore, we should conclude that, fortunately, brain matters too. Very good examples of how the latter is true are the two current ATST and EST projects aiming at a close approximation to the described ideal situation.

\acknowledgements
We have to thank the organizers of the workshop for having given us the opportunity for preparing this contribution. Thanks are due as well to A. V\"ogler and to J.A. Bonet who kindly lend us their simulations and software, respectively. We are also indebted to L.R. Bellot Rubio who prepared Fig.\ \ref{fig:filamento}. This work has been partially funded by the Spanish Mi\-nis\-terio de Educaci\'on y Ciencia, through Projects  ESP2006-13030-C06-02, AYA2009-14105-C06-06, and Junta de Andaluc\'{\i}a, through Project P07-TEP-2687, including a percentage from European FEDER funds.

\end{document}